\documentclass[amssymb,onecolumn,aps]{revtex4}

\usepackage{graphicx}
\usepackage{amsmath, xparse}
\usepackage{color}
\usepackage{hyperref}
\usepackage{subfig}

\usepackage{soul}

\begin{document}
\preprint{APS/123-QED}
\title{Consensus formation on heterogeneous networks }

\author{Edoardo Fadda}
\affiliation{Department of Mathematical Sciences, Politecnico di Torino, \\ISIRES}

\author{Junda He}
\affiliation{Information Systems and Technology Cluster,
School of Computing and Information Systems,\\Singapore Management University}

\author{Claudio J. Tessone}
\affiliation{Blockchain \& Distributed Ledger Technologies, UZH Blockchain Center,\\ University of Zurich}

\author{Paolo Barucca}
 \email{E-mail: p.barucca@ucl.ac.uk}
\affiliation{Department of Computer Science, University College London}

\date{\today}

\begin{abstract}
Reaching consensus -a macroscopic state where  the system constituents display the same microscopic state- is a necessity in multiple complex socio-technical and techno-economic systems: their correct functioning ultimately depends on it. 
In many distributed systems -of which blockchain-based applications are a paradigmatic example- the process of consensus formation is crucial not only for the emergence of a leading majority but for the very functioning of the system. 
We build a minimalistic network model of consensus formation on blockchain systems for quantifying how central nodes - with respect  to their average distance to others - can leverage on their position to obtain competitive advantage in the consensus process. 
We show that in a wide range of network topologies, the probability of forming a majority can significantly increase depending on the centrality of nodes that initiate the spreading. Further, we study the role that network topology plays on the consensus process: we show that central nodes in scale-free networks can win consensus in the network even if they broadcast states significantly later than peripheral ones. 
\end{abstract}

\maketitle
\section*{Introduction}
Blockchains -and other Distributed Ledger Technologies- \cite{Tasca_Tessone_2019, baronchelli2018emergence} staggering success has opened a deep discussion on the possibility of new social, economic and financial  disintermediated paradigms. 
The ambition and the depth of the range of applications of this technology has required an increased level of scrutiny of the underlying mathematical mechanisms, which are still little understood, especially in its consensus formation dynamics. 
The main ingredients of the technology are cryptographic task distribution and coordination. 
The absence of coordination and consensus between users would imply the failure of the protocol to generate a common record of verified data.
In proof-of-work based systems, the possibility of immediately communicating the positive outcome of mining from miners in the distributed network is crucial to minimise the waste of computational power, but on the other hand it serves to maximise the trust in the system \cite{tessone2021stochastic}. There are important (little discussed) assumptions in the consensus of blockchain-based systems, as the analytical derivations rely on negligible propagation times of blocks, and vanishing path-lengths in the peer-to-peer network \cite{nakamoto2009bitcoin}. 

Consensus is the most important trait in the continuously increasing range of blockchain-based systems \cite{spychiger2020tree}. Proof-of-work based open blockchains (of which Bitcoin and Ethereum 1.0 are primary examples) are  decentralised platforms for value exchange through consensus where each participant can contribute to (and is rewarded for) verifying and diffusing the information stored in the common ledger.
The reliability of the information is based on the so-called ``mining'', a computationally-intensive problem-solving task that is  performed independently by each participant; when a participant finds a solution, it shares it with other participating through a peer overlay network. 
Despite the fact that the information communication plays a crucial role in the consensus formation and thus in the functionality of Bitcoin and similar peer networks. 
With the growth of Proof-of-Stake derivatives (and existing protocols such as Stellar and Ripple) \cite{lashkari2021comprehensive} - which all include clock synchronisation algorithms - it is clear that modelling approaches based on discrete time steps has an intrinsic validity in this space. 
With this variety of systems, no systematic studies have quantified the role of the network of communication in shaping the behaviour and the functioning of a decentralised consensus platform. 

In this paper we build a complete theoretical framework for understanding competing consensus in distributed ledger platforms and we establish general results on the emergence of network effects on the system efficiency.  
The main contribution of this study is to model and simulate both the mining activity and the consensus formation and show how closeness centrality of nodes can lead to a competitive advantage in winning consensus.
The role of network centrality affects the strategies of users in the system and can compromise the correct functioning of cryptoeconomies, especially in presence of bandwidth and computational power concentration among a small subset of users. 
In the following section we discuss how researchers have investigated the role of network propagation in blockchain technology. 
In the model section we introduce the notation and simple formalism that allows us to model both the mining activity and the consensus formation mechanism.
In the results section we present experiments to test the competitive advantage of central nodes in different networks.
Finally, we outline possible perspectives on how to extend the model and how to correctly monitor the efficiency and reliability of the networked social systems. 

\section{Literature review}
The role of communication in the functioning of blockchain protocols has been investigated since the very beginning of Bitcoin. 
In \cite{nakamoto2009bitcoin}, as Nakamoto proposed, Bitcoin is enforced with a proof-of-work mechanism. 
By adjusting the mining difficulty, it limits the block generation rate to around 1 per 10 minutes. 
According to \cite{nakamoto2009bitcoin}, if two blocks are received simultaneously, honest miners should always work on their first received block before the occurrence of next block, as miners should always work on the longest chain.  
In fact, miners can deviate from this behaviour, e.g. by deciding to mine new blocks for shorter chains. 
\cite{Du2017Review} produced a systematic review of the blockchain consensus algorithms. 
The work analysed limitations of the proof-of-work in Bitcoin. 
One of the limitations is that increasing mining difficulty encourages the formation of mining pools. 
As a result of forming mining pools, the computational power can become centralised. 
The paper \cite{decker2013information} evaluated the propagation mechanism on updating the ledger replicas. 
They have focused on analysing the relationship between the propagation speed and the blockchain fork rate. 
By experimentally modulating the Bitcoin protocol, they have shown that speeding up the dissemination of information adequately reduce the number of forks.   
Generally, if a node receives a block which conflicts with its previous ledge replica, it can be ignored.
\cite{decker2013information} showed that - due to the information propagation times - a node may find more convenient not to pass on the information of new block, in order to increase the probability of mining an alternative one. 
For honest miners, a fork is an undesired side-effect of the blockchain protocol.  
In \cite{stifter2019merged}, the authors summarise that the fork rate is correlated to the performance, capacity, security level and degree of decentralisation of the blockchain network. 
Different adversarial attacks, such as the celebrated \textit{selfish-mining} \cite{eyal2014majority}, resort on the strategic decision of the nodes as to \textit{when} share the newly created block. Empirical analyses \cite{li2020proof,li2020mining}, and modelling approaches \cite{schwarz2021agent}  have found indications that such attacks take place in smaller-sized blockchains.
Moreover, the fork rate over a period could be considered as a highly relevant indicator of the resources utilisation level and malicious behaviours (e.g. selfish mining) for a particular time period. 
Based on the orphaned blocks dataset produced by Blockchain.info, a downtrend on the number of orphaned blocks is observed during the past few years. 
In \cite{till2019fork}, the authors suggested that this drop of the number of forks indicates an improvement of the block propagation mechanism in Bitcoin. 
They produced an empirical analysis of forks in the Bitcoin network and concluded that the probability of the earlier propagated block to be included in the main chain increases linearly with the time advantage it gained over the competing block.
Both \cite{decker2013information} and \cite{till2019fork} mention that the propagation time of different blocks varies.  
The overall propagation delay is constituted by the transmission time and verification time of the block. 
From \cite{decker2013information}'s measurements, the propagation delay varies drastically for blocks that are smaller than 20KB meanwhile the delay costs tend to be constant for blocks larger than 20KB. 
In \cite{croman2016scaling} the statistics of block propagation time is reported in a general evaluation of the problem of scaling of cryptoeconomies, i.e. the possibility of the decentralised services to keep working when the number of users increase. 
Recently, \cite{shahsavari2019performance,shahsavaritheoretical} showed in a series of simulations that the block propagation delay has a positive linear relationship with the block-size and that increasing the number of neighbours and the bandwidth can significantly speed the block propagation in the network. 
In \cite{shahsavaritheoretical} the authors also evaluate the probability of fork formation and the correspondences to the block size, average P2P bandwidth, and average number of neighbours per node respectively, based on the Erd\H{o}s-R\'enyi random graph model. 

\section{The model of consensus propagation}

In this Section we present the framework that we use for the numerical experiments as well as for the analytical calculations.

\paragraph*{The model}
We here define the network model for discrete-time consensus formation. 
We consider a graph $G = (V, E)$, where $V$ is the set of nodes of the network. Nodes represent  \textit{miners}  (i.e. validator nodes that may produce blocks), each node $i$ is characterised by a computational power $c_i$ and storing a local copy of the blockchain of a given height $h_i$. 
In the following, for convenience, we use the terms miners and nodes interchangeably depending on whether we are focusing on their mining activity or on their network properties. 
$E$ is the set of edges and $t_{ij}\in \mathbb{N}$ is the time it takes for a direct flow of information to go from node $i$ to node $j$. 
In the weighted model, $t_{ij}$ is discrete and at each iteration miner $j$ receives the information on the block that miner $i$ had $t_{ij}$ steps before. 

The propagation time between two generic miners is given by the weighted shortest path between them in the graph, so that the distribution of propagation times \cite{decker2013information} is shaped by the weighted network distances \cite{blondel2007distance}. 
In the following we use binary values for $t_{ij}$ so that at each time step, each miner receives the blocks from its first neighbours in the network. 
Nodes are further endowed with a height $h_i$ which is the length of the local copy of the blockchain they store. 

For each miner, we sample the time for mining the next block. 
The times $t_i$ are sampled from an exponential distribution characterised by a rate parameter $\lambda_i=\lambda \, c_i$ and rounded to the nearest integer above.
Each iteration, the mining times are synchronously updated, $t_i \rightarrow t_i - 1$. 
Miners find the next block when their mining time goes to zero. 
When a block $b$ is mined by a node $i$, it gets appended to node $i$'s local copy of the blockchain, and the node gets assigned a height $h_i \rightarrow h_i + 1$. 
It is worth noting that in this model forks can appear and two or more blockchains with the same or with different heights can propagate at the same time within the network. 
When a block arrives to a node $j$, it can perform one of the following actions:

\begin{itemize}
    \item 
    if the block has a height greater than its blockchain copy, $h_i$, then the node will add it to its blockchain and start mining the next block
    \item
    if the block has a height less than the its blockchain copy then the node will ignore it
    \item
    if two blocks with a height greater than its blockchain copy arrive together, then the node will choose to add the block that is more frequent in its neighbourhood. If both have the same frequency then one at random is selected. 
\end{itemize}

Mining times are updated either when a node finds a new block or when it accepts a new block from a neighbouring node.
This interplay between propagation and mining gives an advantage to the nodes that - due to their position in the network - tend to receive new blocks before others.
With this model it is possible to simulate chaotic behaviour of distributed networks in which the typical mining time is far smaller than the typical propagation time on the network. 

\paragraph*{The probability of forks}
Empirical studies on blockchain protocols \cite{saad2019overview,Geier2019,zora193491} have shown the emergence of forks, i.e. the simultaneous existence of multiple - potentially conflicting - blockchains among different groups of miners. 
In this paragraph we aim at giving a simple statistical intuition of this occurrence and its relationship to the interplay of two characteristic times: the mining times - governed by mining protocols and CPUs - and propagation times - governed by block sizes and network bandwidth. 
In general, the possibility of competing blocks depends on the joint distribution of the order statistics for the smallest mining times, and their relative time gaps. 
The larger the time gaps between the smallest times, the less probable is for late blocks to be able to compete in the consensus formation via the network propagation.
For the sake of simplicity, here and in the following sections, each miner is characterised by the same computational power, $c_i = 1$. 
The parameter of the exponential distributions $\lambda$ can be chosen, depending on the number of miners, so that the expected minimum time is fixed to $10$ minutes.
The probability distribution function for the mining time of each miner is 
\begin{equation}\label{eq:pdf}
p(t) = \lambda \text{e}^{-\lambda t}.
\end{equation}
We are interested in the corresponding distribution for the minimum time, 
\begin{equation}\label{eq:pdfmin}
p_1(t) = N p(t)C(t)^{N-1}= N\lambda \text{e}^{-N \lambda t}
\end{equation}
where $C(t)$ is the cumulative distribution function of the mining time of each individual miner. 
The expected minimum mining time, i.e. the mining time of the whole system, is then
\begin{equation}\label{eq:min}
\langle t_{min} \rangle = \frac{1}{N\lambda}.
\end{equation}
Let us now focus on the joint distribution of the two smallest mining times in our model. 
Supposing an initial situation where all miners start mining at the same time, it reads
\begin{equation}\label{eq:jointmin}
p(t,t') = N(N-1)p(t)p(t')C(t')^{N-2}\theta(t'-t)
\end{equation}
where $\theta(x)$ is the Heaviside step function, so that the distribution of the gap $\Delta = t'- t$ reads:
\begin{equation}\label{eq:delta}
p(\Delta) = \int dt dt' \delta(\Delta - t'+ t)p(t,t') = \lambda(N-1)e^{-\lambda(N-1)\Delta}
\end{equation}
which yields an average $\langle\Delta\rangle = 1/\lambda (N-1)$. A more general expression for the case of heterogeneous miners can be found in the appendix. 
Moreover, this also implies a cumulative probability $C(\Delta)$ of a time difference between the best two times smaller than $\Delta$:
\begin{equation}\label{eq:cdfdelta}
C(\Delta) \approx \lambda (N-1) \Delta
\end{equation}
Based on \cite{decker2013information}, $\Delta$ can be calibrated based on the empirical distribution of propagation times in real blockchain systems, e.g. the median for Bitcoin was $8.7$ seconds in 2015.
For $\Delta$ equal to $8.7$ seconds, and substituting $\lambda (N-1)\approx 1/600$ seconds, then the probability is 
\begin{equation}\label{eq:cdfapprox}
C(\Delta=8.7s) \approx 8.7s/(600s) \cong 0.0145 = 1.45\%
\end{equation}
which is compatible with the old observed value of $1.69\%$ in \cite{decker2013information}, but less so with the more recent estimate of $0.41\%$ \cite{gervais2016security}. 
Despite the simplifying assumptions, the order statistics of a set of exponential distributions provides a numerical agreement with the observed statistics of forks in Bitcoin. \\
The average inter-block time of 10 minutes is on the upper limit of the existing blockchains. Ethereum 1.0, for example, has an average inter-block time of 14 seconds. 
For Ethereum, where the propagation time and the block creation time are comparable \cite{tessone2021stochastic}, we need to compute directly the cumulative distribution function to get the probability of having competing blocks in the system, estimated by the $6.8 \% $ daily ratio between uncle blocks - rewarded, yet unused, mined blocks - and mined blocks \cite{linkystats}. This uncle ratio is roughly compatible - in this simple model - with an estimated $C(\Delta)$ ranging between $3.51\%$ and $5.22\%$. See Table \ref{tab:propagation_times} for similar estimates for other blockchain systems. These estimates do not necessarily reflect real stale blocks rate, potentially outlining the limits of the assumptions of our simple model with homogeneous and independent miners.

\begin{table}[]
\begin{tabular}{|c|c|c|c|c|}
\hline
\textbf{Blockchain} & \textbf{Block creation time (s)} & \textbf{Propagation time (s)} & \textbf{Stale blocks rate} & \textbf{C$(\Delta)$} \\ \hline
BTC       & 600     & 8.7    & 0.41\%        & 1.44\%              \\ \hline
ETH       & 14   & 0.5-0.75        & 6.8\%         & 3.51-5.22\%               \\ \hline
LTC       & 150    & 1.02      & 0.273\%          & 0.68\%                \\ \hline
DOGE       & 60     & 0.85     & 0.619\%      & 1.40\%               \\ \hline
\end{tabular}
\caption{Estimated $C(\Delta)$ from block creation times and median propagation times for multiple blockchain systems, compared with the stale blocks rate. Data are taken from \cite{gervais2016security}. \label{tab:propagation_times}}
\end{table}

\section{Results}
In our experiments we investigate three ensembles of networks: Erd\H{o}s-R\'enyi, stochastic block models (SBM), and Barab\'asi-Albert. 
Erd\H{o}s-R\'enyi provides a solid baseline for evaluating the consensus formation process \cite{barabasi2016network}. 
The comparison with the SBM, where nodes are divided into communities provides a scenario to evaluate the potential emergence of persistent forks in different communities. 
The analysis on Barab\'asi-Albert networks tests a scenario of heterogeneously connected miners in which the distribution of distances has a broader support and node centrality can create great differences between miners' ability to propagate their mined blocks. 
Firstly to monitor the difference in propagation and cluster formation in these networks we compute the distribution of sizes for the clusters of miners keeping the two different new blocks. 
Secondly, we demonstrate through simulations how network centrality can statistically favour some nodes over others in the case of a competition between blocks during the consensus formation. 
The average distance of a miner to other nodes determines which block will prevail in a temporary fork, closeness centrality quantifies this, as it is defined as the reciprocal of the sum of the distances of a node towards all the other nodes \cite{Boccaletti2006}.

We consider graphs with $|V|=1000$, this dimension is big enough to achieve stable results for the proposed experiments and allow us to run several repetitions. In particular, we consider Barab\'asi-Albert graphs characterised by an average degree $\bar{d}=8$, Erd\H{o}s R\'{e}nyi graphs characterised by an edge probability $p_{\mathcal{E}} = 8/1000$, and SBM graphs characterised by four blocks each one containing the same amount of nodes (i.e. $p = [0.25, 0.25,0.25,0.25]$) and the following connection matrix:
\begin{equation}
C =
\begin{bmatrix}
5 & 1 & 1 & 1 \\
1 & 5 & 1 & 1 \\
1 & 1 & 5 & 1 \\
1 & 1 & 1 & 5
\end{bmatrix},   
\end{equation}
where each entry $C_{ab}$ represents the expected number of edges from a node in block $a$ to nodes in block $b$. 
These choices are done in order to set the average degree to be $8$, as in the bitcoin network. Technically, the bitcoin protocol allows to a larger number of connections, nevertheless the default parameters called \textit{max connections} is set to $4$ \cite{Park2019}. The characteristics of the networks are reported in Table \ref{tab:network_characteristics}.

\begin{table}[]
\begin{tabular}{|c|c|c|c|c|}
\hline
\textbf{Graph Type} & \textbf{Diameter} & \textbf{Betweenness} & \textbf{Closeness Centrality} \\ \hline
Erdos          & 6 (0.00)          & 2.58e-3 (2e-5)       & 2.81e-1 (1e-2)                \\ \hline
SBM            & 6 (0.00)          & 2.03e-3 (3e-5)       & 2.71e-1 (1e-2)                \\ \hline
Barabasi       & 6 (0.00)          & 2.23e-3 (1e-5)       & 3.12e-1 (1e-2)                \\ \hline
\end{tabular}
\caption{Main Characteristic of the generated graphs. For each value is reported average and std. dev. in brackets. \label{tab:network_characteristics}}
\end{table}

We present the results of the experiments in the two following subsections.

\begin{figure}
    \subfloat[Erdos graphs with $|V|=1000$]{\label{fig:Erdos_cs}\includegraphics[scale=0.4]{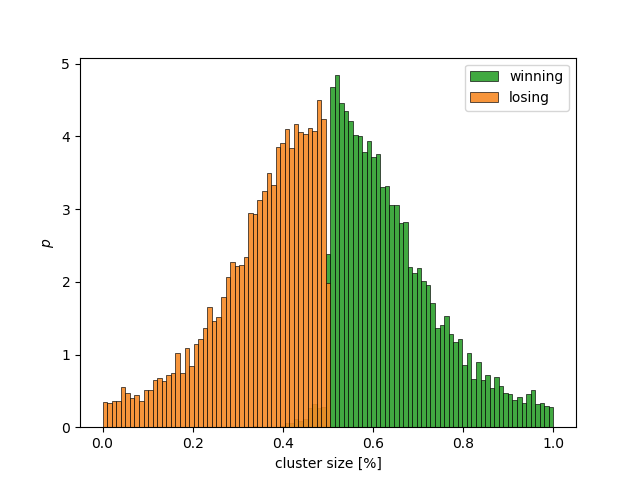}} \\
    \subfloat[SBM graphs with $|V|=1000$]{\label{fig:stoch_block_cs}\includegraphics[scale=0.4]{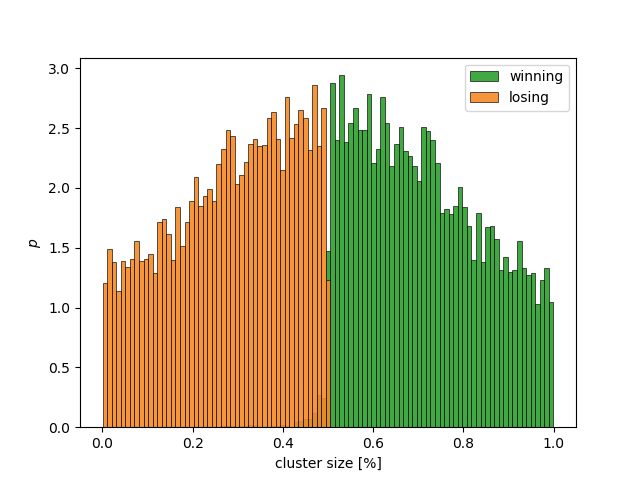}} \\
    \subfloat[Barab\'asi-Albert graphs with $|V|=1000$]{\label{fig:albert_cs}\includegraphics[scale=0.4]{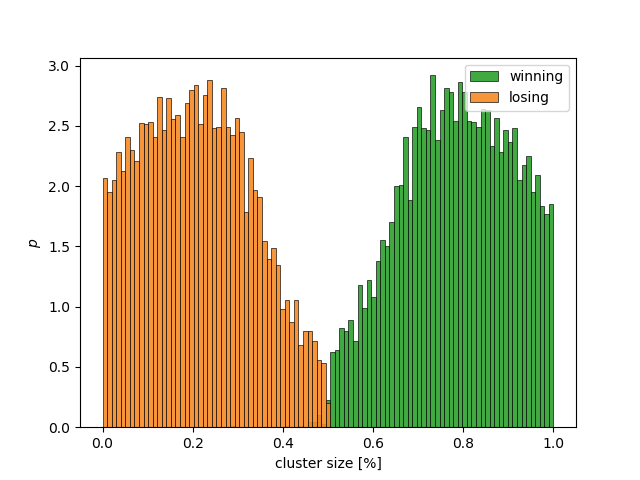}} 
    \caption{Distribution of cluster dimensions at time $t^*=4$ for the different networks}
    \label{fig:cs}
\end{figure}

\paragraph*{Cluster dimension}
The dimension of the winning cluster is one of the most important features of the competitive diffusion process. 
Due to its dynamics, the process ends with the winning cluster having all the nodes and the losing one having no nodes, so to observe differences we need to look at clusters before convergence.
Given a graph with average degree $\bar{d}$, we monitor the sizes of the two clusters for

$$t^*= \left \lceil{\frac{\ln|V|}{\ln\bar{d}} }\right \rceil = \left \lceil{\frac{\ln 1000}{\ln 8 }}\right \rceil = 4 $$
where $\left \lceil{.}\right \rceil $ rounds to the nearest higher integer. 
This is an estimate for the propagation time that is obtained by considering a graph with a tree structure with a fixed number of branches $\bar{d}$. The histograms of the frequency of each cluster size are shown in Figure \ref{fig:Erdos_cs} \ref{fig:stoch_block_cs}, and \ref{fig:albert_cs} for the Erd\H{o}s-R\'enyi, SBM, and Barab\'asi-Albert graph, respectively.

Each graph shows the histogram of the various cluster sizes at a given time, expressed as percentage of nodes. 
The histograms are not strictly symmetric since at time $t^*$, the propagation process is not finished and some nodes do not belong to either of the clusters.
In the central region the two histograms present an overlap. This is due to the cases in which the two clusters have similar size and the one with a lower number of nodes eventually wins the propagation. 
In the Erd\H{o}s-R\'enyi graphs, at time $t^*$ the propagation process has not yet a clear winner, thus the dimension of the cluster are more likely to be similar. 
In the stochastic block model, the situation is similar, but it is characterised by a greater variance. 
Due to the different density of edges, it is possible for a cluster to spread across more than one block, thus gaining in few iterations a big percentage of the network. 
Finally, in the Barab\'asi-Albert graphs the diffusion process at time $t^*$ is close to the end: the winning cluster has a size close to the whole network and the losing one has nearly no nodes. 

\begin{figure}
    \subfloat[Erd\H{o}s network with $|V|=1000$]{\label{fig:erdos_cc}\includegraphics[scale=0.4]{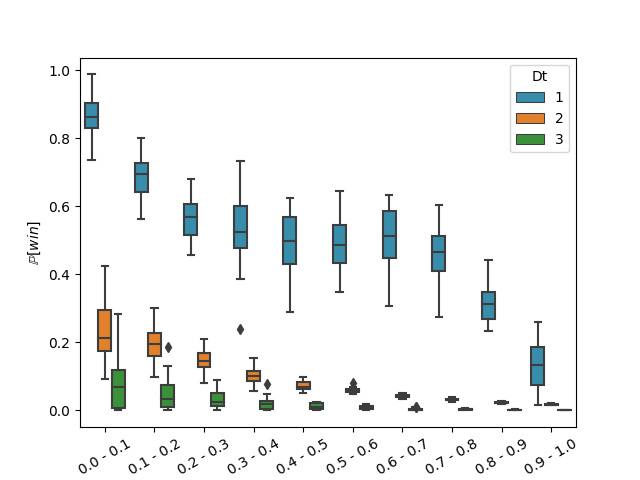}} \\ 
    \subfloat[Stochastic Block network with $|V|=1000$]{\label{fig:stoch_block_cc}\includegraphics[scale=0.4]{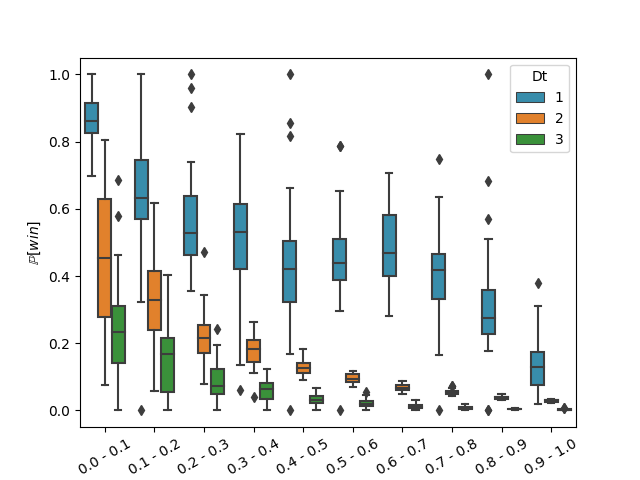}} \\ 
    \subfloat[Barab\'asi-Albert network with $|V|=1000$]{\label{fig:albert_cc}\includegraphics[scale=0.4]{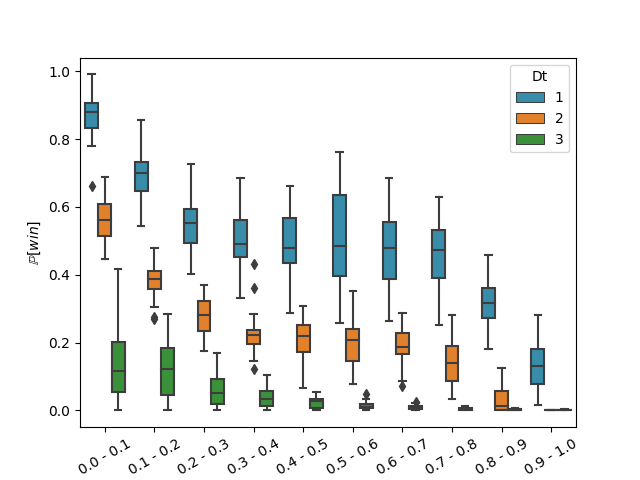}}  
     \caption{Probability of winning for different values of closeness centrality quantile $q$ and different difference in time $\Delta t$ in different graphs.}
    \label{fig:cc}
\end{figure}

\paragraph{Closeness centrality}
In the second experiment we consider the frequencies of victories for a node characterised by a given centrality (measured in terms of quantiles for the network). 
In order to observe this relation, first we sample a graph and we rank nodes according to their closeness centrality. The distributions of closeness centralities can be found in the appendix.
For each quantile of closeness centrality $q$ we select the corresponding node and we run a competitive diffusion process between this node and another randomly sampled node, repeating this competition for $100$ times. 
Each point in the graph represents the empirical probability of winning the process of competitive diffusion for the selected node.
Finally, we average this empirical probabilities over $10$ different graphs for each ensemble and obtain the results shown in Figure \ref{fig:erdos_cc} \ref{fig:stoch_block_cc}, and \ref{fig:albert_cc}, for the Erd\H{o}s-R\'enyi, SBM, and Barab\'asi-Albert graph, respectively. 
All the graphs present the results for $\Delta t = 0$, $\Delta t = 1$ and $\Delta t = 2$ since for $\Delta t = 3$ the graphs reduces to a constant zero line. 
The lower quantiles in the plot correspond to the higher values of closeness centrality. 
As expected, if a node has an high centrality it is more likely to win the competitive diffusion process. Moreover, the time is of crucial importance since, if $\Delta t = 1$ (i.e., the considered node starts the diffusion process one time step after the other) the probability to win suddenly decreases. 
All the curves for all the ensembles, for $\Delta t = 0$, present a plateau around $0.5$ corresponding to the median closeness centrality values.
This is due to the fact that for closeness centrality values closed to the median, the sampled nodes considered to test the victory probability of a node will be, with equal probability, both above and below the closeness centrality value of the node being tested. 

For $\Delta t=1$ and $\Delta t=2$, the results are also different for each of the graph ensembles considered. 
For Erd\H{o}s-R\'enyi, the difference between the curve with $\Delta t=1$ and $\Delta t=2$ is not so big. 
The Erd\H{o}s-R\'enyi model also shows low variance, i.e. the points tend to be closer to their average values.  
SBM graphs display similar curves but present an higher variance. 
Remarkably, for this ensemble also nodes corresponding to low quantiles are able to reach winning probability closed to one. 
These nodes are the ones located near to the intersection of two blocks of the network.
They are characterised by a relatively low degree but they are able to diffuse their cluster in two blocks in a short time. 
Finally, the Barab\'asi-Albert graphs behave quite differently.
The variance is lower than for the SBM graphs but greater than the Erd\H{o}s-R\'enyi graph. 
Despite the advantage of the first node, for $\Delta t = 1$, the second node has still non-negligible probability to win even for relatively low quantiles of closeness centrality. 
This is due to the wider distribution of closeness centrality that characterises Barab\'asi-Albert graphs. 
In fact, such variations increase the advantage of being a node with relatively high closeness centrality. 
\section{Conclusions}
In this study we have shown how information processing and propagation on the heterogeneous overlay network of a distributed system can generate asymmetries in the contribution of system elements to its functioning. 
We have introduced a minimalistic  time-discrete model which can be used to investigate detailed scenarios for the role of networks in selfish and non-cooperative behavior in distributed systems.
The numerical results here presented also address the question on how to identify network features that enable to predict which nodes, or cluster of nodes, will win a competitive propagation process before the actual full propagation takes place. 

This paper opens up a new stream of research with multiple applications to distributed systems, but specifically to blockchains and distributed ledger technologies. How these systems reach consensus is a key differentiating component in their design and concept. While this paper is focused on proof-of-work, the modelling approach is applicable to other scenarios. Newer platforms - like those based in variants of delegated-proof-of-stake, voting models (such as Stellar and Ripple) or others more unique - such as IOTA - depend on round-based systems for their functioning. In those, a discrete-time approach is even more central to uncover possible adversarial attacks that may hinder the emergence of trust. 

\newpage
\section{Appendix A: Order statistics for heterogeneous miners}
When miners have different computational power, then their distribution of mining times changes. 
The probability distribution function for the mining time of any miner is 
\begin{equation}\label{eq:pdfhet}
p_i(t) = \lambda_i e^{-\lambda_i t}
\end{equation}
where $\lambda_i$ can be modelled as proportional to the computational power, i.e. $\lambda_i = \lambda \, c_i$, measured in a common scale $\lambda$.
We are interested in the corresponding distribution for the minimum time, $t_{min} = \min_i\{t_m^i\}$
\begin{equation}\label{eq:pdfminhet}
p_1(t) = \sum_{i=1} p_i(t)\prod_{j\neq i}^NC_j(t)
\end{equation}
where $C_i(t)$ are cumulative distribution functions for the mining times of different miners.
which for the exponential distributions simply reads 
\begin{equation}\label{eq:pdfminhet}
p_1(t) = N \overline{\lambda} e^{- N \overline{\lambda} t}
\end{equation}
where $\overline{\lambda} = \frac{1}{N}\sum_{i=1} \lambda_i$.
The expected minimum mining time, i.e. the mining time of the system, is
\begin{equation}\label{eq:minhet}
\langle t_{min} \rangle =  \frac{1}{\sum_{i}\lambda_i}.
\end{equation}
Let us now focus on the joint distribution of the two smallest mining times in the heterogeneous case. 
Supposing again an initial situation where all miners start mining at the same time, it reads
\begin{equation}\label{eq:cdfhet}
p(t,t') = \sum_{i \neq j} p_i(t)p_j(t') \prod_{k\neq i,j}^N C_k(t')\theta(t'-t)
\end{equation}
which for the exponential distributions reads 
\begin{equation}\label{eq:cdfhetIII}
p(t,t') = \sum_{i\neq j}  \lambda_i \lambda_j e^{-\lambda_i (t-t') - \sum_{k}\lambda_k t'} \theta(t'-t).
\end{equation}
Finally leading to the following equation for the time gap distribution
\begin{equation}\label{eq:deltahet}
p(\Delta) = \frac{ \sum_{i\neq j}  \lambda_i \lambda_j e^{-\sum_{k\neq i}\lambda_k \Delta}}{N\overline{\lambda}},
\end{equation}
which shows the explicit dependence of the time gaps between new mined blocks - potentially leading to forks -  and the distribution of computational power across miners.

\section{Appendix B: Closeness Centrality Distributions}
\begin{figure}[!h]
\includegraphics[scale=0.5]{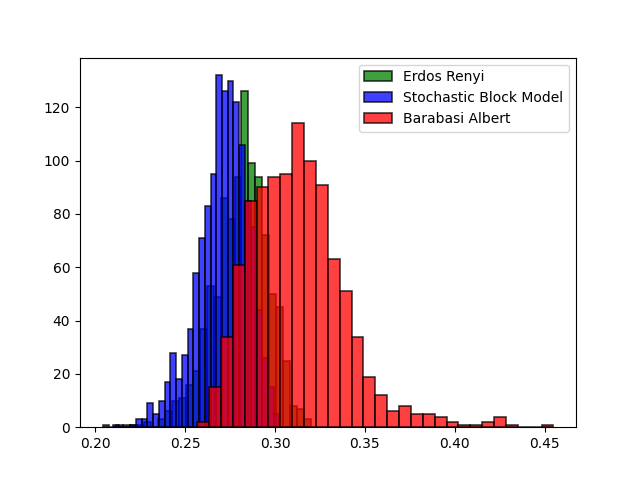}
\caption{Distribution of closeness centrality with $|V|=1000$}
\label{fig:cc_hist_tot}
\end{figure}

Closeness centrality distributions for the 3 topologies is represented in Figure \ref{fig:cc_hist_tot}.

\section*{Availability of data and material}
The full simulation code is available on the  online repository: \hyperlink{https://gitlab.com/edo.fadda/blockchainsimulator}{gitlab.com/edo.fadda/blockchainsimulator}

\section*{Competing interests}
The authors declare that they have no competing interests.
  
\section*{Funding}
PB acknowledges support from the UCL Centre for Blockchain Technologies.  CJT acknowledges funding of SNSF (Switzerland) through project  200021\_182659.

\section*{Authors' contributions}
EF, PB, CJT designed the experiment. EF, JH performed all the simulations in the study. JH performed a preliminary analysis for the study. EF, PB, CJT wrote and edited the final manuscript. 

\section*{Acknowledgements}
The authors would like to thank Dr Paolo Tasca and Kristelle Feghali for insightful discussions.

\bibliographystyle{plain} 
\bibliography{consensus}

\section*{Figures}

\end{document}